\documentclass[cits]{POS}
\usepackage{amsmath}
\usepackage{units}

\renewcommand{\d}{\text{d}}

\title{JEWEL -- a Monte Carlo Model for Jet Quenching}

\ShortTitle{JEWEL -- a Monte Carlo Model for Jet Quenching}

\author{\speaker{Korinna Zapp} \\%

        Physikalisches Institut, Universit\"at Heidelberg, Philosophenweg 12,
		D-69120 Heidelberg, Germany\\

	  ExtreMe Matter Institute EMMI, GSI Helmholtzzentrum
f\"ur Schwerionenforschung GmbH, Planckstra\ss e 1, 64291 Darmstadt, Germany\\ 

        E-mail: \email{zapp@physi.uni-heidelberg.de}}

\author{Johanna Stachel\\

        Physikalisches Institut, Universit\"at Heidelberg, Philosophenweg 12,
		D-69120 Heidelberg, Germany\\

        E-mail: \email{stachel@physi.uni-heidelberg.de}}

\author{Urs Achim Wiedemann\\

        Physics Department, Theory Unit, CERN, CH-1211 Gen\`eve 23,
Switzerland\\

        E-mail: \email{Urs.Wiedemann@cern.ch}}

\abstract{
The Monte Carlo model \textsc{Jewel}~1.0 (Jet Evolution With Energy Loss)
simulates parton shower evolution in the presence of a dense QCD medium. In its
current form medium interactions are modelled as elastic scattering based on
perturbative matrix elements and a simple prescription for medium induced gluon
radiation. The parton shower is interfaced with a hadronisation model. In the
absence of medium effects \textsc{Jewel} is shown to reproduce jet measurements
at LEP. The collisional energy loss is consistent with analytic calculations,
but with \textsc{Jewel} we can go a step further and characterise also
jet-induced modifications of the medium. Elastic and inelastic medium
interactions are shown to lead to distinctive modifications of the jet
fragmentation pattern, which should allow to experimentally distinguish between
collisional and radiative energy loss mechanisms. In these proceedings the main
\textsc{Jewel} results are summarised and a Monte Carlo algorithm is
outlined that allows to
include the Landau-Pomerantschuk-Migdal effect in probabilistic frameworks.
}

\FullConference{4th international workshop High-pT physics at LHC 09\\

                 February 4-7, 2009\\

                 Prague, Czech Republic}

\begin{document}

\section{Introduction}

In ultra-relativistic heavy ion collisions, the produced QCD matter reduces
significantly the
energy of high transverse momentum partons. While most
of the experimental evidence comes from studying the leading hadronic fragments
of the parent partons via single inclusive hadron spectra and jet-like particle
correlations at RHIC, there is ample motivation for studying medium-modified
jets beyond their leading fragments, in particular: i)~At the LHC a larger
fraction of the entire
medium-modified jet
fragmentation pattern will become accessible above background. ii)~Studying the
distribution of subleading fragments is likely to discriminate between different
microscopic
mechanisms conjectured to underly jet quenching, thereby helping to characterise
more
precisely the properties of matter tested by jet quenching. iii)~Modelling the
distribution
of subleading jet fragments is essential for any operational procedure aiming at
disentangling jets from background or characterising the jet-induced
modification
of
the background. Such reasons motivate the development of tools which account
dynamically
for the interaction between jet and medium, and which model medium-modified
jets on the level of multi-particle final states. 

The Monte Carlo technique provides a powerful tool for the simulation of
multi-particle final states. In the presence of medium effects, we expect that a
parton shower can contribute to understanding `jet quenching', in particular for
the following reasons: i)~It reproduces the unmodified jet evolution as
vacuum baseline. ii)~Energy and momentum can be conserved exactly at each
vertex. iii)~ Different microscopic mechanisms for the interaction
between projectile and target can be tested. 
iv)~The modelling of realistic multi-hadron final states is clearly beneficial
for comparing theory and data.

In developing \textsc{Jewel} (Jet Evolution With Energy Loss), our main focus
was to arrive at a code
which allows to study in detail the dynamics relating the evolution of the
parton shower to the microscopic modelling of the medium interactions.

This paper aims at summarising the most important features and results of
\textsc{Jewel}~1.0. For more details and more elaborate discussion we refer
to~\cite{Zapp:2008gi,thesis}.

\section{Vacuum Baseline}

In the absence of medium effects the \textsc{Jewel} parton shower is closely
related to the mass-ordered shower in
\textsc{Pythia}~6.4~\cite{Sjostrand:2006za}. It generates fragmentation patterns
of a parton with
energy $E$, produced in a hard scattering, into a multi-parton final state by
probabilistic iteration of the $1\to 2$ splitting processes. The virtuality
$Q^2$ is used as ordering variable, since it is directly related to the parton
lifetime. The probability that no splitting occurs between an initial and final
virtuality $Q_\text{i}^2$ and $Q_\text{f}^2$, respectively, is described by the
Sudakov form factor
\begin{equation}
\label{eq_sudavac}
 S_{\text a}(Q_{\rm i}^2,Q_{\text f}^2) =
 \exp \left[ - \int \limits_{Q_{\text f}^2}^{Q_{\text i}^2} \!
     \frac{{\text d} Q'^2}{Q'^2} \!\!\int \limits_{z_-(Q'^2,E)}^{z_+(Q'^2,E)}
\!\!\!\!\!\!\!
     {\text d} z \,
     \frac{\alpha_{\text s}}{2\pi}\, \sum_{\text b,c}
     \hat P_{{\text a}\to {\text bc}}(z) \right]\, .
\end{equation}
Here, $\hat P_{\text{a}\to\text{bc}}(z)$ are the standard LO parton splitting
functions for quarks and gluons ($a,b,c \in \{q,g\}$) and $z$ is the energy
fraction carried by the first daughter. The probability density for a splitting
to occur at $Q^2$ is given by 
\begin{equation}
 \Sigma_{\text a}(Q_{\text i}^2,Q^2) 
      = \frac{{\text d} S_{\text a}(Q_{\text i}^2,Q^2)}{{\text d} (\ln Q^2)}
      = S_{\text a}(Q_{\text i}^2,Q^2) \sum_{\text b,c}
        W_{{\text a}\to {\text bc}}(Q^2)\, ,
        \label{eq_dersuda}
\end{equation}
where
\begin{equation}
 W_{{\text a}\to{\text bc}}(Q^2)
       =  \int \limits_{z_-(Q^2,E)}^{z_+(Q^2,E)}\!\!\!\!\!
     {\text d} z \, \frac{\alpha_{\text s}(z(1-z)Q^2)}{2\pi}
     \hat P_{{\text a}\to{\text bc}}(z) \, .
\end{equation}

One thus has to choose the parent parton's virtuality according to
(\ref{eq_dersuda}). The type of splitting is selected consistent with $W_{{\text
a}\to{\text bc}}$ and the energy sharing is given by the splitting function
$\hat P_{\text{a}\to\text{bc}}(z)$. This procedure is repeated for the
daughters until no partons above the infrared cut-off scale $Q_0^2$ are left.
Angular ordering is enforced by allowing only for splittings with decreasing
emission angle.

\smallskip

At $Q_0$ the parton shower is interfaced with a hadronisation model, that has
 to be flexible enough to be usable for the complex parton state in heavy-ion 
collisions. In \textsc{Jewel} a variant of the Lund string
fragmentation~\cite{Andersson:1983ia} is used, where the knowledge about the
colour flow is
replaced by the assumption of maximal colour correlation of partons close in
momentum space.

\medskip

\begin{figure}[t]
\centering
\input{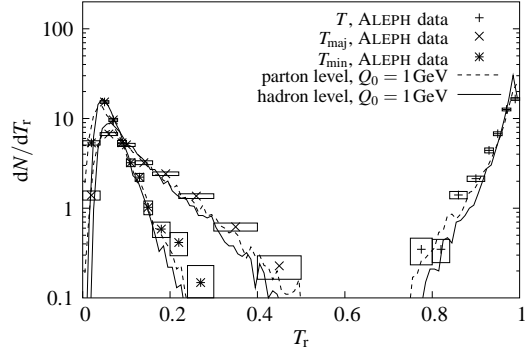}
\caption{The thrust, thrust major and thrust minor
($T_\text{r}=(T,T_\text{maj},T_\text{min})$) distributions for
$\sqrt{s}=\unit[200]{GeV}$ $e^+e^-\to q\, \bar{q}\to X$
collisions. Data of the ALEPH Collaboration~\cite{Heister:2003aj} 
are compared to simulations of \textsc{Jewel}. Dashed line: parton level after
parton
shower evolved down to $Q_0 = \unit[1]{GeV}$; solid line: hadron level
after parton
shower evolution followed by hadronisation ($Q_0 = \unit[1]{GeV}$). Figure
taken from~\cite{Zapp:2008gi}.}
\label{fig_thrustvac}
\end{figure}

The \textsc{Jewel} parton shower is validated against a set of benchmark
measurements in $\sqrt{s}=\unit[200]{GeV}$ $e^+e^-$ collisions at LEP. In
Fig.~\ref{fig_thrustvac} we compare the results of the simulation for the event
shape variables thrust $T$, thrust major $T_\text{maj}$ and thrust minor
$T_\text{min}$ to data, with the definitions,
\begin{equation}
T \equiv {\rm max}_{\vec{n}_T} 
	\frac{\sum_i \vert \vec{p}_i\cdot \vec{n}_T\vert}{\sum_i \vert
	\vec{p}_i\vert} \, .
\qquad 	
T_{\rm maj} \equiv {\rm max}_{\vec{n}_T\cdot \vec{n}=0} 
	\frac{\sum_i \vert \vec{p}_i\cdot \vec{n}\vert}{\sum_i \vert
	\vec{p}_i\vert}
\qquad
T_{\rm min} \equiv 
	\frac{\sum_i \vert \vec{p}_{ix}\vert}{\sum_i \vert \vec{p}_i\vert}
\end{equation}
Thrust is a measure for how pencil-like an event is. For events, in which
all
momenta are (anti)parallel to the thrust axis $\vec{n}_T$, $T=1$ and for
spherical events $T=1/2$. Repeating the analysis in the plane transverse to the
thrust
axis one obtains thrust major. 
Thrust minor sums up the components $\vec{p}_{ix}$ of the final particle
momenta $\vec{p}_{i}$, which are  orthogonal to the plane defined by
$\vec{n}$ and $\vec{n}_T$.

As seen in Fig.~\ref{fig_thrustvac}, the final state parton shower provides a
reasonable description of these jet event shapes over most of the measured
range. The \textsc{Jewel} parton shower is not matched to exact matrix elements,
which may be the reason why it gives too few events with large $1-T$,
$T_\text{maj}$ and $T_\text{min}$. Fig.~\ref{fig_thrustvac} also shows that the
thrust variables, since they are infrared and collinear safe quantities, are not
very sensitive to hadronisation.

The comparison to other event shape variables is of similar or
better quality~\cite{thesis}, which shows that \textsc{Jewel} accounts with a
sufficient accuracy for global features of
the momentum flow.

In comparison to jet event shapes, there are measurements which are more
sensitive to the discrete and stochastic nature of the partonic processes
underlying the QCD jet fragmentation. One such measurement is the $n$-jet rate,
which is based on $k_\perp$-clustering algorithms and quantifies the
sub-structure of jets. The overall agreement of the simulation results with
data is reasonably good, although the 3-jet rate is somewhat underestimated
(possibly due to the missing matrix element matching).

\begin{figure}[t]
\centering
\input{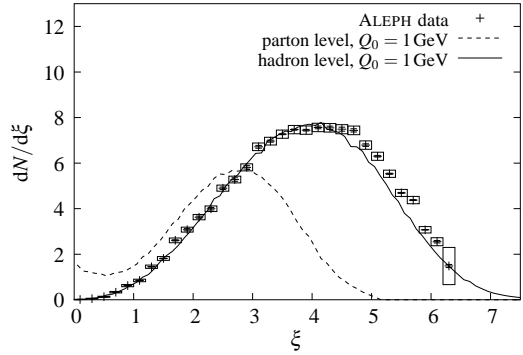}
\caption{The inclusive distribution $\d N_{\rm ch}/\d \xi$ with $\xi = \ln
\left[ E_\text{jet}/p_\text{hadron} \right]$ of charged hadrons in $e^+e^-\to
q \bar q \to X$ events at $\sqrt{s}=\unit[200]{GeV}$.
Data of the ALEPH Collaboration~\cite{Heister:2003aj} 
are compared to simulations of \textsc{Jewel}. Dashed line: parton level after
parton
shower evolved down to $Q_0 = \unit[1]{GeV}$; solid line: hadron level after
parton
shower evolution to $Q_0 = \unit[1]{GeV}$ followed by hadronisation. Figure
taken from~\cite{Zapp:2008gi}.}
\label{fig_xivac}
\end{figure}

In contrast to the measurements discussed so far, the modelling of 
single inclusive intra-jet hadron distributions and multi-hadron correlations
requires detailed knowledge about the hadronisation mechanism. This is seen for
instance in Fig.~\ref{fig_xivac}, where we compare results of our simulation to
data for the inclusive distribution $\d N_\text{ch}/\d \xi$, with
$\xi=\ln(E_\text{jet}/p_\text{hadron}$. Irrespective of the choice of $Q_0$
the partonic distribution is very hard while hadronisation is expected to soften
the distribution considerably. As seen in Fig.~\ref{fig_xivac}, \textsc{Jewel}
with the modified string fragmentation describes the data to a level better
than a few percent.

We note that in the presence of a high-multiplicity environment, novel
hadronisation mechanisms may play a role. Within these uncertainties, which are
mainly related to the modelling of hadronisation, we have established that
\textsc{Jewel} provides a reliable baseline for the characterisation of jet
quenching phenomena.

\section{Medium Modification of Jets}

We regard the medium as a collection of partons acting as scattering centres.
For the case of elastic interactions between the jet and the medium, each
scattering centre displays to the partonic projectiles an elastic $2\to 2$
scattering cross section $\d\sigma/\d t$. For this exploratory study the medium
is modelled as an ideal gas of quarks and gluons with a constant temperature
$T$. The masses of the scattering centres are fixed to
$m_\text{scatt}=\mu_\text{D}(T)/\sqrt{2}$, where $\mu_\text{D}(T)$ is the
thermal Debye mass.

To specify the spatiotemporal structure of the parton shower, we start from
the
estimate that the parton shower evolves down to components of virtuality $Q_f$ 
on a time scale $1/Q_f$. For a parton of energy $E$ and mass $Q_f$, this
will
be
time-dilated in the rest frame of the medium to about $E/Q_f^2$. 
If the parton originated from the branching of some parton of virtuality $Q_i$,
then
the parton of virtuality $Q_f$ existed  for a duration of approximately
\begin{equation}
	\tau = \frac{E}{Q_f^2} - \frac{E}{Q_i^2}\, .
\end{equation}
For the parent parton, which initialised the parton shower, the lifetime is
$\tau =  \frac{E}{Q_i^2}$. In the case of a medium of  constant density $n$
 scatterings occur during this time with the probability
\begin{equation}
	1 - S_{\rm no\,  scatt}(\tau) = 1 - \exp\left[  - \sigma_{\rm elas}\, n\,
\tau \beta
\right]\, .
\end{equation}
For the differential elastic scattering cross section, we choose the LO
perturbative $t$-channel parton-parton cross section, which gives the dominant
contribution.
To make contact with the previous studies, we have chosen two
different regularisation schemes for the elastic scattering cross section,
namely case~I
\begin{equation}
\label{eq_sigmai}
	\sigma^{\rm elas} = \int \limits_0^{\vert t_{\rm max}\vert} d\vert
t\vert\,
	\frac{\pi\, \alpha_s^2(\vert t\vert + \mu_D^2)}{s^2}
	C_{R}\, \frac{s^2 + (s-\vert t\vert)^2}{\left(\vert t\vert + \mu_D^2
\right)^2}\, ,
\end{equation}
which is the default, and case~II
\begin{equation}
\label{eq_sigmaii}
	\sigma^{\rm elas} = \int \limits_{\mu_D^2}^{\vert t_{\rm max}\vert} d\vert
t\vert\,
	\frac{\pi\, \alpha_s^2(\vert t\vert )}{s^2}
	C_{R}\, \frac{s^2 + (s-\vert t\vert)^2}{\vert t\vert ^2}\, .
\end{equation}

Elastic scattering may occur between subsequent splitting processes and after
the last splitting, provided the parton is still inside the medium.

The model introduced here does not yet include a mechanism of radiative energy
loss. Since this may be the main source of energy degradation, we have included
an option to enhance the vacuum splitting functions by a factor $(1 + f_{\rm
med})$,
\begin{equation}
	\hat{P}_{a\to bc}(z) \longrightarrow (1 + f_{\rm med})\, \hat{P}_{a\to
bc}(z)\, ,
\end{equation}
as long as the splitting occurs in the medium. This prescription has been argued
to display characteristics of radiative energy loss~\cite{Borghini:2005em}. 

It is assumed that the hadronisation of a jet is unaffected by the medium,
i.e.\, occurs after all splittings and medium interactions. Since it is not
clear, how jets in nuclear collisions hadronise, we investigate two options.
One is that only the hadronisation of the parton shower is considered, whereas
in the second the parton
shower and the recoiling scattering centres hadronise together. 

\medskip

\begin{figure}[t]
\centering
\input{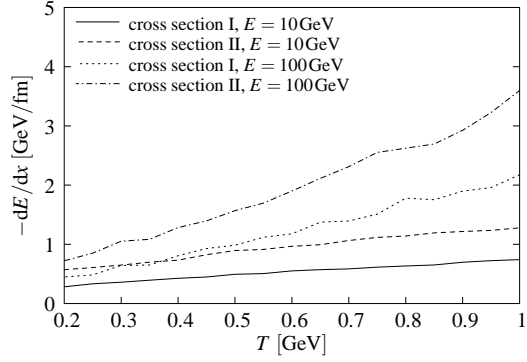}
\caption{The average parton energy loss $\d E/\d x$ of a quark of energy $E$,
undergoing multiple elastic collisions over a path length $L = \unit[1]{fm}$
in a thermal medium of temperature $T$. 
Elastic collisions are described by the infrared regulated partonic cross
sections case~I and case~II. Figure
taken from~\cite{Zapp:2008gi}.}
\label{fig_dedx}
\end{figure}

In Fig.~\ref{fig_dedx}, we have calculated the resulting average energy loss
for
an in-medium path length of $L = \unit[1]{fm}$ in a medium of temperature
$T$.
Quantitatively, the
models I and II show differences of approximately a factor 2 in $\d E/\d
x$ for a \unit[10]{GeV} parton. Cross section~II leads to a larger energy
loss, 
as may be expected since there is minimum momentum transfer. We find that the
temperature dependence of $\d E/\d x$ shown in Fig.~\ref{fig_dedx},
is consistent with the dependences reported
previously~\cite{Bjorken:1982tu,Thoma:1990fm,Braaten:1991we,Djordjevic:2006tw,
Adil:2006ei,Zakharov:2007pj,Peigne:2008nd,Domdey:2008gp,Zapp:2005kt}. The
differences between
cases~I and~II are representative of the typical factor 2 uncertainties
between
different model studies, though some recent studies lead to slightly larger
values
of  $\d E/\d x$ than those shown in Fig.~\ref{fig_dedx}, see
e.g.~\cite{Peigne:2008nd}.

\smallskip

\begin{figure*}[t]
\centering
\input{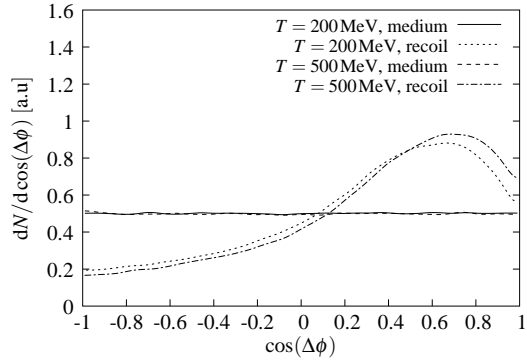}
\caption{
Angle with
respect to jet axis of
recoiling scattering centres as compared to the undisturbed medium for
different temperatures (cross
section~I, $E_\text{jet}=\unit[100]{GeV}$, $L=\unit[1]{fm}$). Hadronisation is
not included but may affect these distributions significantly. Figure
taken from~\cite{Zapp:2008gi}.}
\label{fig_bckgrnd}
\end{figure*}

The energy lost by a jet is redistributed amongst the target components.
Characterising the recoil of the medium may provide a means to disentangle
different mechanisms of parton energy loss. In Fig.~\ref{fig_bckgrnd} the 
angular distribution of recoiling scattering centres is compared to the
undisturbed thermal medium. The recoil moves predominantly in the jet direction
with a maximum at $\Delta \phi \sim 0.8$ nearly independent of the
temperature. At face value, Fig.~\ref{fig_bckgrnd} indicates that a jet can be
accompanied by
additional multiplicity which has its maximum separated from the jet axis
by a characteristic finite angle $\Delta\phi$. We note, however, that whether
the partonic distributions of Fig.~\ref{fig_bckgrnd} will or will
not
change
substantially upon hadronisation may depend on details of the hadronisation 
mechanism. 

\smallskip

\begin{figure*}[t]
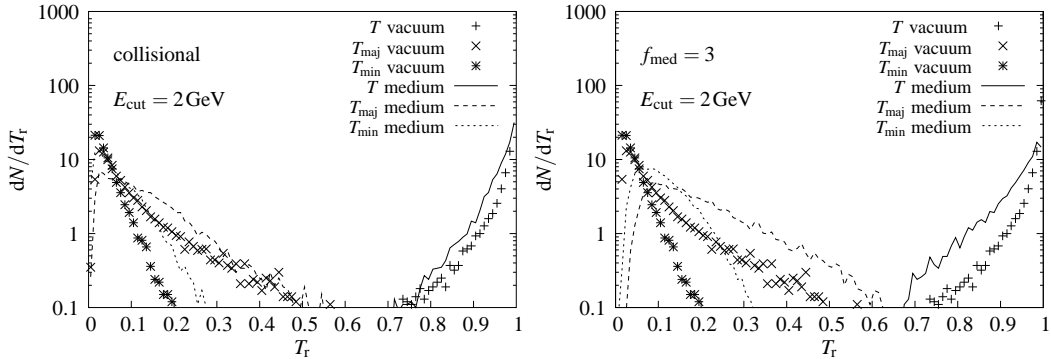

\centering
\input{9c-thrustmed-lqcd300.pstex_t}
\input{9d-thrustmed-lqcd300.pstex_t}
\caption{Thrust, thrust major and thrust minor
($T_\text{r}=(T,T_\text{maj},T_\text{min})$) for a single
\unit[100]{GeV} jet.
The \textsc{Jewel} parton shower in the vacuum is compared to
two scenarios including medium-induced parton energy loss. Left hand side:
Collisional
energy loss for a medium of $T=\unit[500]{MeV}$ and in-medium path length
$L=\unit[5]{fm}$ (the recoil is hadronised with the medium).
Right hand side: Radiative energy loss for $f_{\rm med} = 3$ and
$L=\unit[5]{fm}$. Only hadrons with energy 
above $E_{\rm cut} = \unit[2]{GeV}$ are included. Figure
taken from~\cite{Zapp:2008gi}.}
\label{fig_thrustmed}
\end{figure*}

Next, we study to what extent jet measurements are sensitive to medium effects.
Only hadrons with energies above a soft background cut are included in the
event shape analyses. This may indicate the extent to which different jet
modifications remain visible above the hadronic background in heavy-ion
collisions. Fig.~\ref{fig_thrustmed} shows the thrust distributions for a
single jet with and without medium modifications. Collisional energy loss is
seen to lead to only a mild broadening of the distributions. The kinematics of
elastic $2\to 2$ scattering dictates that more energetic projectiles are
deflected by smaller angles. 
The recoiling scattering centres have mostly relatively low
momenta so that the medium-induced broadening becomes small if only hadrons of
energy above $E_{\rm
cut}=\unit[2]{GeV}$ are included in the analysis.
In contrast to collisional energy loss, single partonic components of
medium-induced
additional radiation have a higher probability to carry a significant energy
fraction of
the initial projectile energy. So, on general grounds, one expects that the
medium-induced
broadening of the distributions  in thrust, thrust major and thrust minor will
persist even
if soft hadrons of energy $E_\text{h} < E_{\rm cut} = \unit[2]{GeV}$ are dropped
from the analysis.
This is clearly seen in Fig.~\ref{fig_thrustmed}\,(right). The jet rates exhibit
the
same overall behaviour. Measurements of event shape variables and jet rates may
thus provide a tool for disentangling collisional and radiative energy loss.

\begin{figure*}[t]
\centering
\input{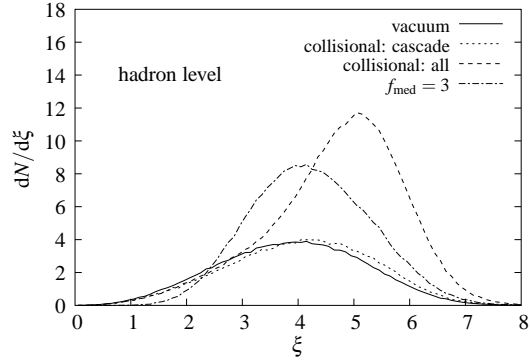}
\caption{The single inclusive distribution $\d N/\d\xi$ for a single
medium-modified quark jet ($E_q = \unit[100]{GeV}$) after hadronisation ($Q_0 =
\unit[1]{GeV}$) (only charged hadrons are included). Collisional energy loss is
calculated 
for $T = \unit[500]{MeV}$ and $L = \unit[5]{fm}$, 
with recoil partons either hadronised together with the cascade
('all') or not included in the hadronisation ('cascade').
Medium induced radiation is calculated for $f_{\rm med} = 3$ and $L =
\unit[5]{fm}$. Figure
taken from~\cite{Zapp:2008gi}.}
\label{fig_ximed}
\end{figure*}

In Fig.~\ref{fig_ximed}, we plot the medium modification of the single
inclusive
distribution
$\d N/\d\xi$ in the presence of collisional and radiative medium effects.
 If elastic interactions with the medium are
included but only the cascade is hadronised, then
the total jet multiplicity of projectile hadrons 
depends only very weakly on the
medium, since $2\to 2$ processes do not increase the
parton multiplicity. However, the total jet multiplicity
may increase significantly if recoil partons are counted towards
the jet. 
Fig.~\ref{fig_ximed} also indicates that radiative mechanisms
result in a strong increase of intra-jet multiplicities.

\section{Outlook}

\begin{figure}[t]
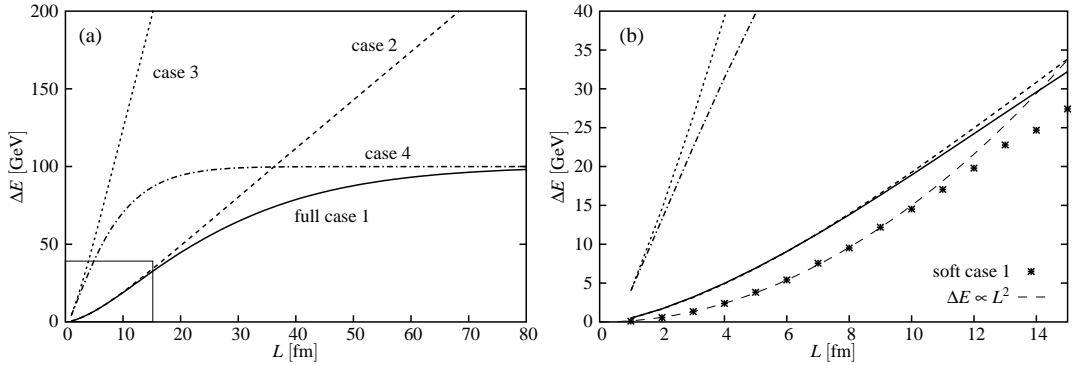

\begin{center}
\input{deltae_vs_l-all-new.pstex_t}
\input{deltae_vs_l-small-new.pstex_t}
\end{center}
\vspace{-0.5cm}
\caption{The medium-induced energy loss $\Delta E$ of a quark of initial
energy $E_q = 100\, {\rm GeV}$ as a function of the in-medium path length $L$.
Results for the full MC algorithm with LPM effect and energy 
conservation (full case 1) are compared to: i) LPM effect without exact energy
conservation (case 2), ii) incoherent limit without energy conservation (case
3), iii)
incoherent limit with energy conservation (case 4), and iv) full case 1 with
constraint 
that momentum transfer per scattering centre is limited to be soft
(soft case 1). The bottom plot zooms into the small $L$-region. In this example
$L_\text{c} \simeq \unit[9]{fm}$.
Figure
taken from~\cite{Zapp:2008af}.}
\label{fig_lpm}
\end{figure}

The most important future development of \textsc{Jewel} is to consistently
include inelastic scattering. In~\cite{Zapp:2008af} a local and probabilistic
implementation of the non-abelian Landau-Pomerantschuk-Migdal effect was
presented. Analytic calculations~\cite{Wiedemann:2000za} suggest that, in
general, the LPM effect can be implemented in a probabilistic Monte Carlo
algorithm by requiring
that the momentum transfer from different scattering centres to the partonic
projectile acts 
totally coherently for gluon production, if it occurs within the formation time
$t_\text{f}=2 \omega/ k_\perp^2$, and that it acts
incoherently, if it occurs after $t_\text{f}$. The Monte Carlo implementation
of this prescription reproduces as a limiting case the BDMPS~\cite{Baier:1996sk}
results without modifying the gluon emission vertex. But it is more flexible in
details of the modelling and has a natural way to implement exact
energy-momentum conservation. In Fig.~\ref{fig_lpm} the average energy loss
for a projectile quark of initial energy $E_\text{q}=\unit[100]{GeV}$
traversing a medium of length $L$ is shown for different cases. Results for the
full MC algorithm, including LPM-effect and exact 
energy-momentum conservation at each vertex are shown as 'full case 1'. Since
the BDMPS multiple soft scattering approximation neglects the high-$q_\perp$
tail of the scattering cross section, we also consider a case where the
momentum transfer is limited to be soft ('soft case 1'). This case shows the
characteristic $L^2$-dependence of the average energy loss. Allowing for hard
scattering, on the other hand, leads to a loss of coherence and thus increases
the energy loss. 'Case 2' is inspired by the high energy approximation assumed
in many analytical studies. In this case the gluon energy has to be smaller than
the initial quark energy, but the total radiated energy in unconstrained. As
expected, this is a good approximation in the region $L<L_\text{c} \sim
\sqrt{4 \omega_\text{max}/\hat q}$, but for
$L>L_\text{c}$ a quasi-incoherent regime is entered, with $\Delta E \propto L$
eventually violating energy conservation. 'Case 3' is a scenario with
incoherent gluon emission without energy conservation (but requiring $\omega <
E_\text{q}$), whereas 'case 4' stands for incoherent gluon radiation with
energy conservation. Energy conservation can be seen to become the dominant
constraint for $L>L_\text{c}$. The difference between cases 1 and 4 illustrates
the quantitative importance of the LPM-effect. 

\medskip

With this tool a dynamically consistent implementation of inelastic scattering
including the LPM-interference in \textsc{Jewel} seems feasible. Among other
planned future extensions are a realistic treatment of the medium and the
generalisation to massive quarks.

\acknowledgments{This research project has been supported via a Marie Curie
Early Stage Research Training Fellowship of the European Community's Sixth
Framework Programme under contract number (MEST-CT-2005-020238-EUROTHEPHY), by
the Helmholtz Alliance
Program of the Helmholtz Association,
contract HA216/EMMI "Extremes of
Density and Temperature: Cosmic Matter
in the Laboratory", and the German BMBF.}

\end{document}